\documentclass[fleqn,usenatbib]{mnras}
\usepackage{graphicx}

\usepackage{newtxtext,newtxmath}
\usepackage{longtable}

\newcommand\xion[2]{#1$\;${\scshape{#2}}}

\newcommand{\dtg}{\ensuremath{{\rm DTG}}}
\newcommand{\dtm}{\ensuremath{{\rm DTM}}}
\newcommand{\deltaX}{\ensuremath{\delta_X}}

\newcommand{\Hmol}{\ensuremath{{\rm H}_2}}

\newcommand{\logNHI}{\ensuremath{\log N(\mbox{\xion{H}{i}})}}

\def\hi{H~{\sc i}}
\newcommand{\nhi}{\ensuremath{N(\mbox{\xion{H}{i}})}}
\newcommand{\nmol}{\ensuremath{N({\rm H}_2)}}

\title[Observed Dust Surface Density]{Observed Dust Surface Density Across Cosmic Times}

\author[P\'eroux, De Cia \& Howk]{C\'eline P\'eroux$^{1,2}$\thanks{E-mail: celine.peroux@gmail.com}, Annalisa De Cia$^{3}$ and J. Christopher Howk$^{4}$
\\
$^{1}$ European Southern Observatory, Karl-Schwarzschildstrasse 2, D-85748 Garching bei M{\"u}nchen, Germany\\
$^{2}$ Aix Marseille Universit\'e, CNRS, LAM (Laboratoire d'Astrophysique de Marseille) UMR 7326, 13388, Marseille, France \\
$^{3}$ Department of Astronomy, University of Geneva, Chemin Pegasi 51, 1290 Versoix, Switzerland\\
$^{4}$ Department of Physics and Astronomy, University of Notre Dame, Notre Dame, Indiana 46556, USA\\
}

\date{Accepted 2023 April 20. Received 2023 April 20; in original form 2022 December 20}
\pubyear{2020}

\begin{document}
\maketitle

\begin{abstract}
Our ability to interpret observations
of galaxies and trace their stellar, gas, and dust content over cosmic time critically relies on our understanding of how the dust abundance and properties vary with
environment. Here, we compute the dust surface density across cosmic times
to put novel constraints on simulations of the build-up of dust. We provide observational estimates of the dust surface density consistently measured through depletion methods across a wide range of environments, going from the Milky Way up to z=5.5 galaxies. These conservative measurements provide complementary estimates to extinction-based observations. In addition, we introduce the dust surface density distribution function -- in analogy with the cold gas column density distribution functions. We fit a power law of the form: $\log f( \Sigma_{\rm Dust})=-1.92 \times \log \Sigma_{\rm Dust} - 3.65$ which proves {slightly} steeper than for neutral gas and metal absorbers. {This observed relation, which can be computed by simulations predicting resolved dust mass functions through 2D projection, provides new constraints on modern dust models.}

\end{abstract}



\section{Introduction}

Dust grains absorb stellar light in the ultraviolet (UV)-optical wavelengths and re-emit it
in the far-infrared, which represents 30\%–50\% of the
radiative output of a galaxy {\citep{RomanDuval17}}. Therefore, our ability to interpret observations
of galaxies and trace their stellar, gas, and dust content with redshift across the entire spectral range critically relies on measurements of the dust abundance and properties
vary with environment and cosmic times. This in turn requires us to understand the
processes responsible for dust formation, destruction, and
transport, as well as their associated timescales.

A fraction of metals in the interstellar medium of
galaxies in both the local and high-redshift Universe resides
in microscopic solid particles or dust grains \citep{Field74,Savage96,jenkins2009, decia2016}. Interstellar dust has manifold impact on the physics and chemistry of the interstellar medium \citep{Zhukovska16, Zhukovska18, Galliano22} as well as the intra-cluster medium \citep{Shchekinov22}. Because dust locks some elements
away from the gas phase, it affects our measurements of the metallicity of galaxies. One of the most important roles of interstellar
grains is that they facilitate the formation of molecular
hydrogen, \Hmol, on their surfaces \citep{Hollenbach71}. The \Hmol\ molecule is the main component of molecular clouds, which are the cradle of star formation in most of the Universe \citep{Klessen16}. Because dust absorbs UV
emission from young massive stars and re-emits it in the
infrared, the spectral energy distribution from dust
is one of the primary indicators of star formation \citep{Calzetti00}.

Yet, in both local and high-redshift galaxies, the dust production rates
in evolved stars \citep{Bladh12, Riebel12} and supernova remnants \citep{Matsuura11, Lesniewska19, Slavin20} are largely insufficient compared to the dust destruction
rates in interstellar shocks \citep{Jones96} to explain
the dust masses of galaxies over cosmic times \citep{Morgan03, Boyer12, Rowlands14,Zhukovska13}. This so-called dust budget crisis
poses an important challenge to our modelling of dust into a 
cosmological context \citep{Mattsson21}. 

Observationally, the amount of dust in astrophysical objects has been quantified with a number of different methods probing various dust signature. {\it Extinction} refers to the amount of light dimming due to all the material lying along the {line of sight between} the astrophysical object and the observer. Extinction is therefore an integrated quantity which encapsulates absorption and scattering away from the line-of-sight. The observational determination of
extinction requires backlights such as stars, Gamma-Ray Bursts, quasars, or other objects
with much smaller angular extent than a galaxy. The extinction at a given wavelength results from
a combination of the grain size distribution \citep{Mattsson20b}, metallicity \citep{Shivaei20} and the optical properties of the grains (which is itself
dependent on the chemical composition of the grains). Therefore, the extinction scales with dust column density or surface density. Similarly, infra-red emission has been used to probe the dust content of galaxies \citep{Chang21}. In particular, significant work has been put into
characterizing these quantities in galaxies beyond the Milky Way, assessing
observational constraints both on the integrated values
\citep[e.g.,][]{remy2014, de-vis2019} and spatially-resolved values within
galaxies \citep[e.g., ][]{vilchez2019}. These works have placed particular
emphasis on the variation of the dust properties with metallicity, stellar
mass, star formation rate, and gas content of the galactic environments, as
these help shape our understanding of the factors that drive the formation/destruction balance of dust. {\it Reddening}, {expressed through the colour excess, E(B--V), }quantifies the differential extinction.

{\it Attenuation} represents the effect of dust on the light continuum from the geometric mix of stars and dust in galaxies. It thus reflects the net effect on light due to a combination of multiple effects, including extinction, scattering back into the line-of-sight as well as contribution from unobscured stars \citep{Salim20}. {Reddening is often expressed in terms of UV continuum slope, $\beta$ \citep{Shivaei20b}. The UV continuum slope depends on the column density of dust along the line-of-sight to
the observer that is dimming the UV light of background objects. A fully consistent comparison between depletion-estimated extinction and
colour-based estimates is found in \cite{Konstantopoulou23}.}

The total far-infrared emission is a proxy for the total dust mass. At z$>$5, that emission is successfully probed at mm wavelengths with facilities such as ALMA. The dust temperature however is less well-constrained in these early times \citep[see Figure 1 of][]{Bouwens20}, leading to a degeneracy in the current estimates \citep{Faisst20, Sommovigo20, Bakx21, Sommovigo22,ChenYungYing22, Fudamoto22, Viero22, Drew22, Ferrara22}. Alternatively, the dust mass is being derived from the mm-continuum of high-redshift galaxies and the gas mass is then estimated assuming a dust-to-gas ratio -- often taken to be the value of the Milky Way \citep[][but see \cite{Popping22}]{Scoville17, Dunne22}. The ratio of dust emission in infrared to the observed UV emission, known as the infrared excess (IRX), is a
measure of the UV dust attenuation. The attenuation/extinction curve/law is characterised by its slope and normalisation \citep[e.g.][]{Calzetti00}. Two galaxies having the same attenuation curves (i.e. same shape) might still differ in their normalisation (i.e. column density of dust). The two quantities, $\beta$ and IRX, are often related into one diagram \citep{Meurer99, Faisst17, Shivaei20b}. The relation is sensitive to a range of interstellar properties including dust geometries, dust-to-gas ratios, dust grain properties, and the spatial distribution of dust. The relation provides a powerful empirical constraint on dust physical properties because it
laid the foundation for a straightforward correction of UV emission in galaxies where infrared observations are lacking (especially at higher redshifts), based on the easily observable UV slope (or colour).

Dust is made of some of the available metals produce by stars. In the interstellar medium, a fraction of these elements is in the gas, and the rest is locked up in dust. Indeed, most metals are under-abundant in the interstellar gas of the Milky Way
\citep{jenkins2009}, reflecting the amount of dust in our galaxy. {\it Dust depletion} can be used to give hints on the dust composition \citep{Savage96, Jenkins14, Dwek16, Mattsson19, RomanDuval22a} and this often indicates that significant amounts of Fe-rich dust should be present in the interstellar medium. Dust measurements based on depletion estimated at UV and optical wavelengths might suffer from a bias where the dustier systems would be obscured at these wavelengths. In that sense, dust depletion provides conservative measurements which can be seen as a lower limit on the total amount of dust in a population.
Depletion measurements provide a direct estimate of the dust content \citep{Dwek98, Draine07, galliano2018}. The depletion can be used to calculate the extinction through the gas which is proportional to the column density of metals \citep[see equation 1 of][]{Savaglio03}. In their Figure 9, \cite{wiseman2017} offer a comparison of the two measurements between depletion-estimated extinction and colour-based estimates in a sample of Gamma-Ray Bursts, indicating significant discrepancies highlighting the possible limitations described above \citep[see also][]{decia2013, zafar2014}.

These multiple observational results have triggered a number of simulation efforts. These works come into two main families: i) semi-analytical models which use empirical relation to approximate some of the physical processes at play; and ii) hydrodynamical cosmological simulations which include full treatments of dust production and destruction. Contemporary models have reached a new level of realism by including a large number of physical processes  \citep{Draine03}. Dust {\it shattering} refers to the breaking of large grains into small dust grains, due to high-velocity collisions. Conversely, {\it coagulation} describes the processes of large grains being made out of smaller entities because of low-velocity collisions. Therefore, both these processes do not change the mass but the size distribution of dust grains. {\it Sputtering} refers to dust destruction by shocks, including those produced by Supernovae blasts.
{\it Astration} reflects the process of dust being absorbed by stars. We note that the same processes can both produce and destroy small/big grains, so that the production sources and destruction sinks are complex processes to simulate. 

A number of efforts based on semi-analytical models have made predictions on the dust mass of galaxies \citep{Bekki15a, Pantoni19, Lapi20, Gjergo20, Dayal22} and associated dust mass function \citep{Popping17,Triani20, Vijayan19}. As subset of these studies has provided information on the dust surface density specifically \citep{Bekki13, Bekki15b, Osman20a, Gjergo20}. In parallel, there have been a number of works introducing dust physical processes within hydrodynamical cosmological models \citep{Moseley22}. Many report estimates of the dust mass density \citep{Gioannini17, Aoyama18, Lewis22}, while others report the dust mass function \citep{Mckinnon17, graziani2019, li2019, hou2019, Baes20}. A number of these efforts have made predictions on dust surface densities in particular, as the ones presented here \citep{McKinnon16, Trayford19}.

The goal of this work is two-fold. On one hand, we provide observational estimates of the dust surface density consistently measured through depletion methods across a wide range of environments, going from the Milky Way up to z=5.5 galaxies. While previous works have estimated the dust mass in local galaxies \citep{de-vis2019, Millard20, DeLooze20, Morselli20, Casasola20, Nanni20, Galliano21}, this study focuses on the dust column density. On the other hand, we introduce the dust surface density distribution function -- in analogy with the gas (\nhi\ or \nmol) column density distribution functions \citep{peroux2003, zwaan2005, zwaan2006, klitsch2019, Peroux20, Szakacs22}. {Spatially resolved simulations predicting the dust-mass function \citep{Pozzi20, Millard20} can predict the dust surface density distribution function through 2D projection. Thus, the observed dust surface density distribution function potentially offers new constraints on modern dust models.}

The manuscript is organized as follows: Section \ref{sec:methods} presents the methods used in this study. Section \ref{sec:results} details how dust surface density relates to the global physical properties of galaxies.
We summarize and conclude in Section \ref{sec:conclusions}. Here, we adopt an H$_0$ = 67.74 km s$^{-1}$ Mpc$^{-1}$, $\Omega_M$ = 0.3089, and $\Omega_{\Lambda}$ = 0.6911 cosmology. We use the latest solar abundance values from \cite{Asplund21}.


\section{Differential Dust Depletion}
\label{sec:methods}

\subsection{Dust Depletion in our Galaxy}

\subsubsection{Methodology} 
\label{sec:methods_MW}

\begin{figure*}
	\includegraphics[width=2.0\columnwidth]{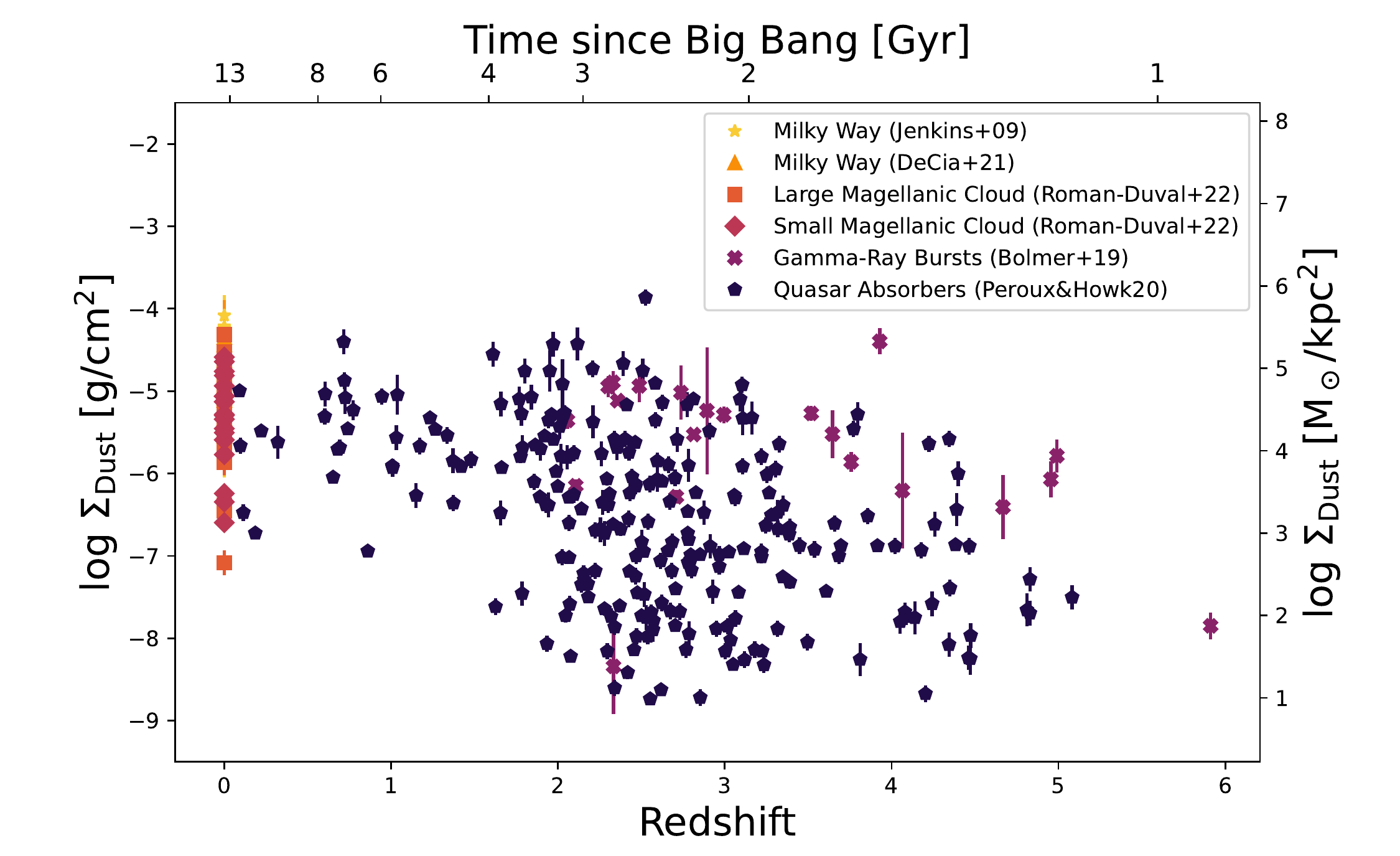}
 	\caption{{\bf Dust surface density as a function of cosmic times.} In this figure and the following, the dust column densities are derived from depletion measurements performed in absorption at UV wavelengths. The shade of colours from light to dark refers to the Milky Way, the Large and Small Magellanic Clouds, Gamma-Ray Bursts and quasar absorbers at z$>$0. We note that Gamma-Ray Bursts galaxy hosts preferentially lie above the quasar absorbers. There is also a clear trend of increasing dust surface density with cosmic time with a large scatter at any given redshift, as expected from the building of dust with time. }
    \label{fig:redshift}
\end{figure*}

The depletion of metals is differential, with some elements showing a higher affinity for incorporation into solid-phase grains than others, based on their chemical properties \citep{Pettini97, vladilo2002, jenkins2009}. The differential nature of elemental depletion has traditionally been used to correct the observed abundance for unseen metals. Early works used lightly-depleted elements to derive abundances (e.g., focusing on Zn or S or to a lesser degree Si). More recent works have taken advantage of the patterns of differential depletion to estimate the extent of the dust-depletion correction. These efforts follow the spirit of \citet{vladilo1998} and \citet{jenkins2009}, using the observed abundances to study the dust depletion beyond the Milky Way  \citep{jenkins2017, RomanDuval22a}.

Specifically, \cite{decia2016} developed a method to characterize dust depletion, \deltaX, without assumption on the total gas+dust metallicity. This is achieved through the study of relative abundances of several metals with different nucleosynthetic and refractory properties, as follows. The relative gas-phase abundance of the metals X and Y is written: $[X/Y] = \rm log (N(X)/N(Y)) - log(N(X)_{\odot}/N(Y)_{\odot})$. This approach enables to calculate the depletion without assumptions on the total metallicity of the gas, including metals locked onto dust grains \citep[see also][]{DeCia21, decia2018a}. {We derive the depletions of different elements as follows:}

\begin{equation}
\deltaX = A2_X + (B2_X \times [Zn/Fe]_{\rm fit}),
\label{eqn:depletion}
\end{equation}

{where [Zn/Fe]$_{\rm fit}$ traces the overall amount of dust depletion and is taken from \cite{DeCia21}. This quantity is equivalent to the observed [Zn/Fe], although it is based on the observations of all available metals.  The coefficients A2$_X$ and B2$_X$ are taken from \cite{Konstantopoulou22}. }

The total dust-corrected metallicity is then computed as:

\begin{equation}
[X/H]_{\rm total} = [X/H]_{\rm observed} - \deltaX
\label{eqn:delta_MW}
\end{equation}

where {[X/H]$_{\rm total}$ is the total metallicity including metals locked into dust grains,} [X/H]$_{\rm observed}$ is the observed abundance of X in the gas phase, and \deltaX, i.e. the logarithm of the fraction of X in the gas phase. Given that each element has a different propensity to deplete onto dust grains, the observations of multiple element ratios provide a measure of the interstellar depletions for various elements X. {Given estimates of $\deltaX$ and observations of [X/H]$_{\rm observed}$, one can derive the quantity: [X/H]$_{\rm total}$.}

\subsubsection{Observational Results in the Milky Way}

In the Milky Way, depletions have been studied for many
heavy elements in several hundreds of sightlines through the diffuse neutral medium \citep{Field74, Phillips82, Jenkins09, DeCia21}. Here, we make use of results from two works: while \cite{jenkins2009} assumes solar metallicity (and solar abundance pattern) for the Milky Way, \cite{DeCia21} report both depletion and metallicity measurements along different line-of-sight. {For consistency with other measurements and to avoid complications related to ionisation corrections, we focus here on log N(H) $\geq$ 20.3 measurements from \cite{jenkins2009}. We also note that the methodology of \cite{jenkins2009} follows the one presented for the Magellanic Clouds next as a fixed global metallicity is assumed.} These studies indicate that in total about 50\% of the metals in the
Milky Way's interstellar medium are incorporated into grains \citep{Draine03}. 
The dust-to-gas (\dtg) and dust-to-metal (\dtm) mass ratios are calculated from these values of the dust depletions (see Section~\ref{sec:results}).

\subsection{Dust Depletion in the Magellanic Clouds}

\subsubsection{Methodology}

The method that we use to estimate the amount of dust in the Magellanic Clouds is slightly different from the one used to estimate the same quantity in the Milky Way. The depletion for element X is expressed as follows:

\begin{equation}
\deltaX = [X/H]_{\rm observed} - [X/H]_{\rm assumed\_total}
\label{eqn:delta_MC}
\end{equation}

where $[X/H]_{\rm assumed\_total} = \log (X/H)_{\rm assumed\_total} - \log (X/H)_{\odot}$ is the total abundance of element X (gas + dust) which here is assumed to
be equal to the abundance of element X in the photospheres of young
stars that have formed out of the interstellar medium \citep[as done for the Milky Way by][]{Savage96, jenkins2009, Tchernyshyov15, jenkins2017, RomanDuval19}. These works compare the chemical abundances in neutral
interstellar gas based on UV spectroscopy to stellar abundances
of OB stars and HII regions \citep[e.g. ][]{Luck98, Trundle07, Hunter07, Toribio17}  to estimate $\deltaX$ from equation~\ref{eqn:delta_MC}. In principle, there could be variations of the metallicity of the neutral interstellar medium in the Magellanic Clouds. 
We stress that the approach described in this section is therefore different from the ones reported in other sections for the Milky Way and for high-redshift galaxies.

\subsubsection{Observational Results in the Large Magellanic Cloud}

The Large Magellanic Cloud lies just 50 kpc from us \citep{Subramanian09}. Its dust content is probed through its extinction map \citep{Furuta22}, while dust reddening \citep{ChenBQ22} and extinction \citep{Gordon03} have been measured in both the Large and Small Magellanic Clouds as well as dust emission \citep{Chastenet17}. Recently, the Large Magellanic Cloud has been the focus of a HST Large Program dubbed "The Metal Evolution, Transport, and Abundance in the Large Magellanic Cloud" (METAL) and introduced in \cite{RomanDuval19}. \cite{RomanDuval21} demonstrates that the depletion of different elements in these data are tightly correlated with the gas (hydrogen) surface density. \cite{RomanDuval22a} further make a new appraisal of the dust estimates in the Milky Way, Large and Small Magellanic Clouds. The Galaxy is more strongly affected by dust depletion than the Large Magellanic Cloud, and even more than the Small Magellanic Cloud. Nevertheless, the way different elements deplete into dust is very similar between these various environments \citep{decia2018a, Konstantopoulou22}.   
Here, we use the data shown in Figure 7 of \cite{RomanDuval22a}, which also include results from \cite{Tchernyshyov15}.
We use a constant metallicity throughout the cloud, taken to be [X/H]=$-$0.30 \citep{RomanDuval22a}. 

 \subsubsection{Observational Results in the Small Magellanic Cloud}

Similarly, depletion of gas-phase metal abundances are clearly seen in the Small Magellanic Cloud, situated at about 60 kpc  \citep{Subramanian09}, though with a smaller degree of depletion reflecting the lower dust-to-metal mass ratio in this system. Initial works from \cite{Tchernyshyov15, jenkins2017}
 provide the first estimates of dust depletion in multiple lines-of-sight. Here, we use the observations displayed in Figure 6 of \cite{RomanDuval22a}.These works assume a constant metallicity throughout the Small Magellanic Cloud, taken to be [X/H]=$-$0.70 \citep{RomanDuval22a}.

\begin{figure*}
	\includegraphics[width=2.0\columnwidth]{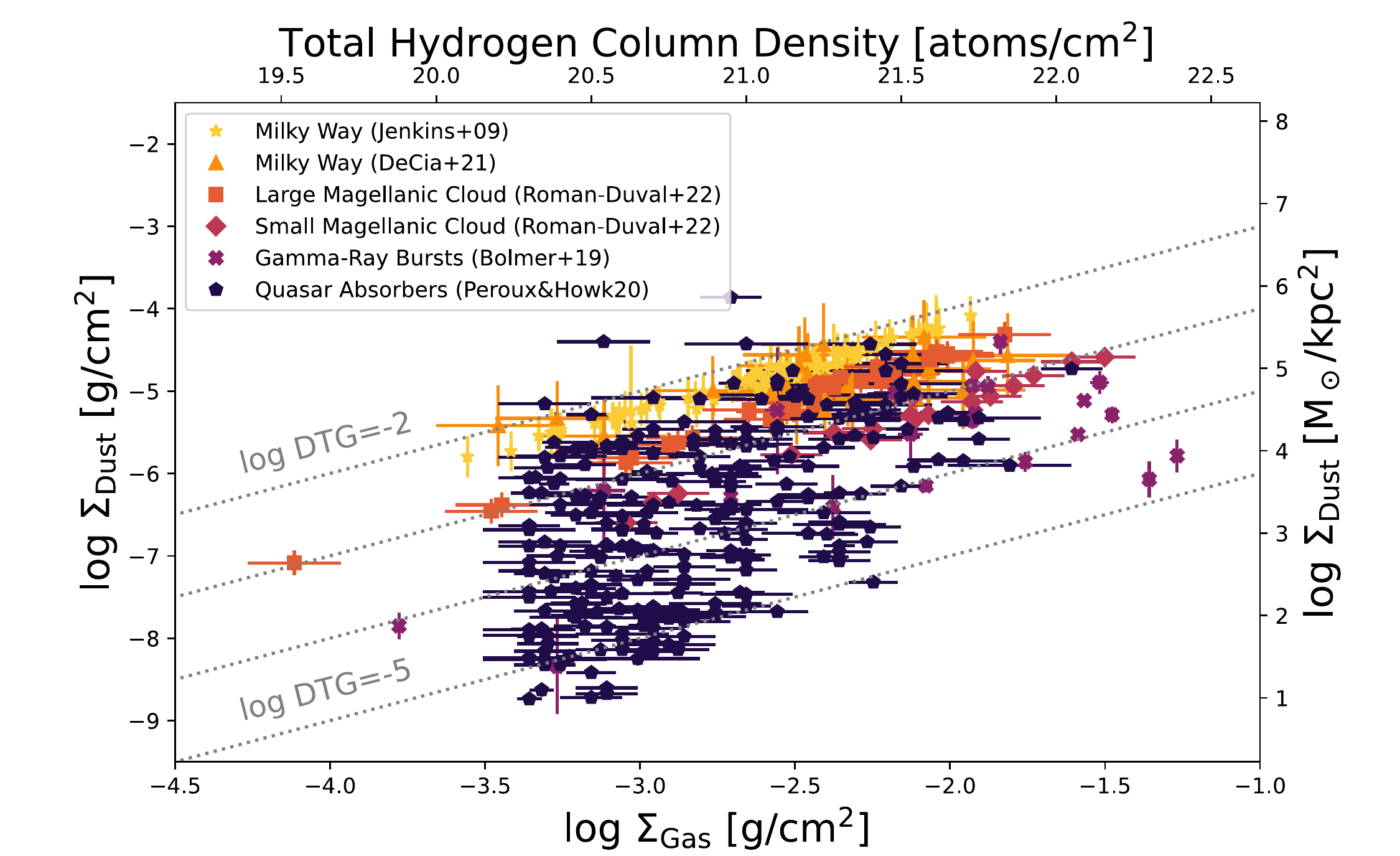}
 	\caption{{\bf Dust surface density as a function of cold gas surface density.} The grey lines represent constant dust-to-gas ratios in steps of 1 dex. The cold gas refers to the sum of neutral atomic, \hi, and molecular, \nmol, except for quasar absorbers where the molecular gas is found to be negligible \citep{ledoux2003}. Overall, the dust column densities follow the total hydrogen column densities with a 4-order of magnitude scatter. }
    \label{fig:gas}
\end{figure*}

\subsection{Dust in High-Redshift Galaxies}

\subsubsection{Methodology}

Given the presence of dust in a broad range of galaxies, depletion is naturally expected 
to be also detected in the material traced by high-redshift absorption lines seen against the background light of bright sources. 
In extragalactic systems, there have been numerous studies of diffuse neutral medium depletions, facilitated by the redshifting of the rest-frame UV absorption lines in the visible range \citep{Ledoux02, vladilo2002, decia2016, Bolmer19}. Similar to the Milky Way, we derive the dust-to-gas ratios for individual absorption systems from the elemental depletions.
Because the behavior of each metal species varies \citep{jenkins2009}, the estimates of the depletion, \deltaX, for each individual element,
$X$, are derived from the dust sequences following the approach highlighted by \citet{decia2016, decia2018, Peroux20}. {The methodology used here is therefore identical to the one described in Section~\ref{sec:methods_MW}.}

\subsubsection{Observational Results in Gamma-Ray Burst Hosts}

Gamma-Ray Burst events in particular probe the insterstellar medium of their galaxy hosts. These systems have been used to probe the dust and metals in Gamma-Ray Burst host galaxies \citep{Savaglio04, Schady07, DeCia12, zafar2019}. 
Here, we report observations from \cite{Bolmer19} which offer a set of dust-depletion estimates based on the correction from \cite{decia2016}, hydrogen column density and metallicity for a sample of Gamma-Ray Bursts.

\subsubsection{Observational Results in Quasar Absorbers}

Quasar absorbers probe the gas inside and around hundreds of foreground galaxies unrelated to the background quasars \citep{pettini1994, Dessauges04, rafelski2012}. There is strong evidence for the differential depletion of metals in the quasar absorbers observed both for the neutral gas
\citep{decia2016, decia2018, Peroux20} as well as for partially-ionized gas \citep{quiret2016, fumagalli2016}. The depletion in quasar absorbers is typically smaller than in the Milky Way \citep{decia2016, RomanDuval22b, Konstantopoulou22, Konstantopoulou23}. In addition, the abundance ratios change in a similar way from local to high-redshift galaxies \citep{DeCia18b, Konstantopoulou22}, indicating that the depletion of dust evolves homogeneously all the way to high-redshift systems. These results also imply that grain growth in the interstellar medium is an important process of dust production

Here, we use the values of dust depletion summarised by \cite{Peroux20}, which are based on results from \cite{decia2018} with some additional updates \citep[see footnote 5 of][]{Peroux20}. We note that only a small fraction (of the order 10\%) of these systems have detections of molecular hydrogen, \nmol. In addition, when detected, the fraction of molecular gas in the cold phase is also found to be small \citep[$\sim$0.01\%, see][]{petitjean00, noterdaeme2008, Balashev19}. For these reasons, we neglect the molecular hydrogen gas in quasar absorbers and assume N(H)=\nhi. 

\begin{figure*}
    \includegraphics[width=2.0\columnwidth]{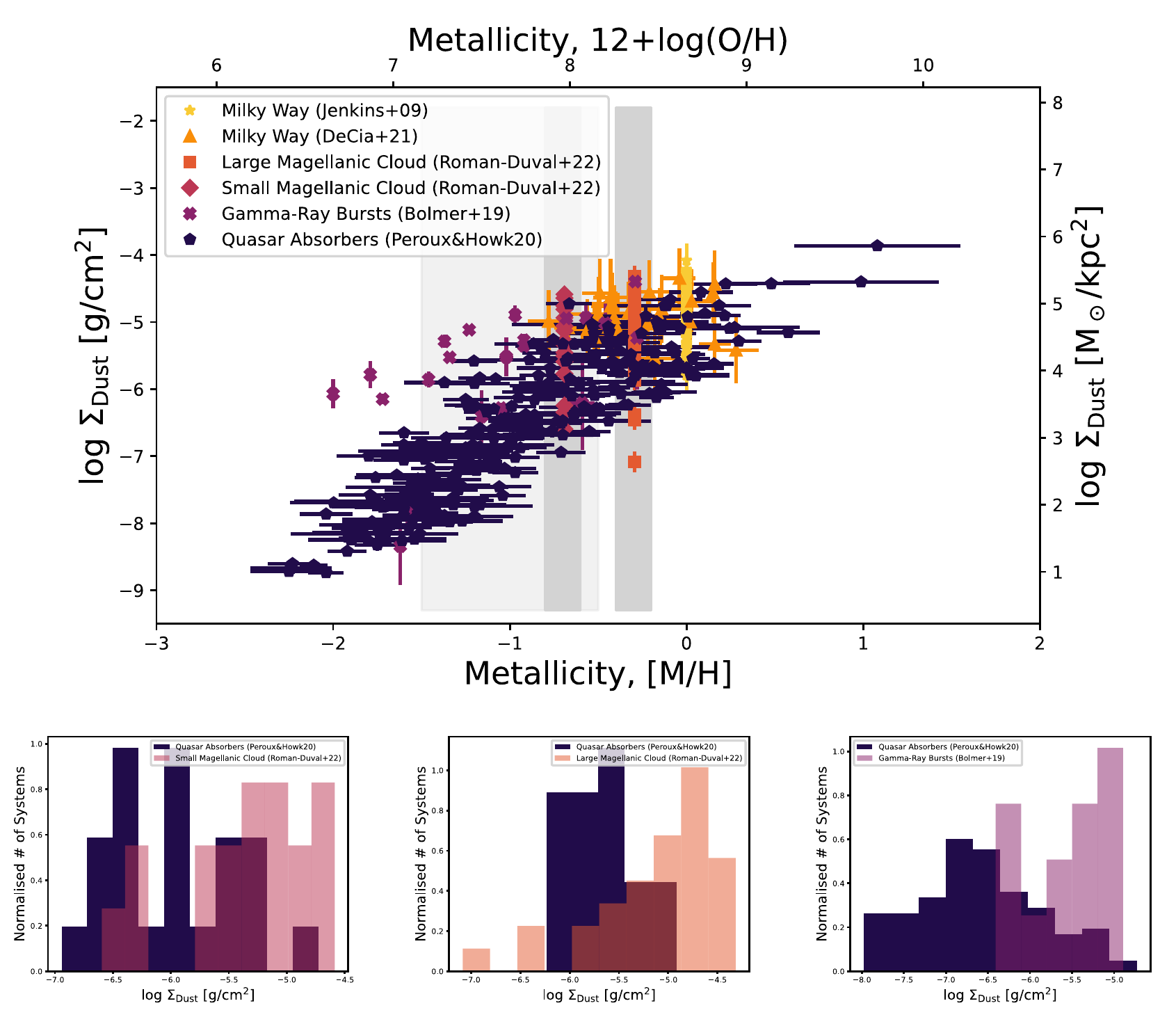}
    \caption{ {\em Top Panel:} {\bf Dust surface density as a function of gas metallicity.} There is a tight correlation between dust surface densities and the dust-corrected metallicity estimates at all cosmic times. For a given metallicity, quasar absorbers have lower surface density of dust than GRB host galaxies and the Milky Way, which are closer to the denser and colder parts of their galaxies. The Milky Way values from Jenkins (2009) and the Large \& Small Magellanic Clouds \citep{RomanDuval22a} are assumed to have one global metallicity each. {\em Bottom Panels:} {\bf Distribution of dust surface density in quasar absorbers, the Magellanic Clouds, and GRB hosts at fixed metallicity.} The bottom left panel displays observations for the SMC, the bottom middle panel shows data for the LMC, while the bottom right whos measurements from GRB hosts. At fixed metallicity, the distribution in dust surface density in the LMC is wider than for quasar absorbers. Similarly, at a given metallicity, $\Sigma_{\rm dust}$ is higher in GRB sight lines that are likely closer to the denser and colder parts of their galaxies than quasar absorbers.     \label{fig:sigvmetal} }
\end{figure*}


\section{Observed Dust Surface Density}
\label{sec:results}

\subsection{Methodology}
The characterisation of the bulk statistical properties of dust involve
assessing the dust-to-gas (DTG) and dust-to-metal (DTM) mass ratios. The former is the fraction of the interstellar mass
locked into dust grains; the latter is the fraction of the
metal mass incorporated into the solid phase. The dust-to-gas ratio for an individual element, $X$, is related to its depletion, $\deltaX$, and its dust-to-metal ratio as follows:
\begin{equation}
\dtg_X =  (1 - 10^{\deltaX}) \, Z_{\rm total}^{\rm X}  = \dtm_X \, Z_{\rm total}^{\rm X}
\label{eqn:dtg}
\end{equation}
where Z$_{\rm total}^{\rm X}$ is the intrinsic abundance of X expressed by mass \citep[e.g.,][]{vladilo2004,decia2016}. We derive the $\deltaX$ from {the differential depletion of various elements as described in Section~\ref{sec:methods}}, and use them to calculate the individual $\dtg_X$ (equation~\ref{eqn:dtg}). 

\begin{equation}
\dtg=\sum_{x}\dtg_x
\label{eqn:dtgsum}
\end{equation}

To obtain a global \dtm\ ratio expressed for all the elements and in mass fraction, we then average the $\dtm_X$ and weight them by elemental abundances and atomic weight for the Milky Way and the Magellanic clouds \citep[as also done in][]{RomanDuval21}. {For the high-redshift galaxies, we derive the global \dtm\ from the \dtg\ as follows:}

\begin{equation}
Z_{\rm total}=\sum_{x} Z_{\rm total}^x
\label{eqn:Z}
\end{equation}
\begin{equation}
\dtm=\dtg/Z_{\rm total}=\frac{\sum_{x}\dtg_x}{\sum_{x} Z_{\rm total}^x}
\label{eqn:dtm}
\end{equation}

For this calculation, we include the 18 elements which depletion have been characterized by \citet{Konstantopoulou22}, though C, O, Si, Mg, and Fe contribute a major fraction of the total dust mass \citep[see also][]{Konstantopoulou23}. In the calculation we also include all the volatile metals that have an element abundance higher than $ 12 + \log(X/H) >$3 \citep[Table 1 of][see also \cite{Asplund21}]{Asplund09}, most notably the N and Ne, which do not contribute to the dust budget, but contribute to the metal budget. 

We then directly calculate the dust surface density which provide a observationally-based measurements of the dust quantity. Specifically, we use several UV-based depletion measurements to calculate the dust mass surface density, $\Sigma_{\rm Dust}$,  in various environments. To this end, we couple observations of the total dust-to-gas ratio, DTG, described above with estimates of the column density of hydrogen gas, both in its atomic and molecular phases. We derive the dust surface densities as:

\begin{equation}
\Sigma_{\rm Dust}=\dtg \times \Sigma_{\rm Gas}=\dtg \times N(H) \times m_{\rm H}\times \mu [g/cm^2]
\label{eqn:DustCol}
\end{equation}

where the total hydrogen column density is N(H)=\nhi+2N(H$_2$), the sum of \nhi\ and two atoms of hydrogen, \nmol, expressed in atoms/cm$^2$. The quantity {$\Sigma_{\rm Gas}$ is the gas surface density}, $m_{\rm H}$ is the hydrogen mass $m_{\rm H}$=$1.67 \times 10^{-24}$ g, and $\mu$ is the mean molecular weight of the gas which is taken to be 1.3
(76\% hydrogen and 24\% helium by mass). The dust surface density, $\Sigma_{\rm Dust}$, is therefore expressed in g/cm$^2$ or alternatively in M$_{\odot}$/kpc$^2$. The resulting values for each of the systems are listed in the table available on-line, an excerpt of which is presented in the Appendix~\ref{app}.

\subsection{Dust Surface Density as a function of Gas Properties}
\label{sec:Prop}

For completeness, we start by briefly summarising the relations between the dust surface densities and galaxy physical properties. This work intentionally refrains to plot dust-to-gas ratio relation with galaxy's properties since those have presented elsewhere \citep[e.g.][]{Popping22}. Figure~\ref{fig:redshift} displays the dust surface density as a function of cosmic time. In this figure and the followings, the left y-axis displays the dust surface density in units of g per cm$^2$ and the right y-axis in units of  M$_{\odot}$/kpc$^2$. We stress that all these dust surface densities are derived from depletion measurements performed in absorption at UV wavelengths. The shade of colours from light to dark refers to the Milky Way, the Large and Small Magellanic Clouds, Gamma-Ray Bursts and quasar absorbers at z$>$0. There is a clear trend of increasing dust surface density with cosmic time with a large scatter at any given redshift, as expected from the build up of metals and associated dust with time. Gamma-Ray Burst host galaxies overall have higher dust surface densities than quasar absorbers at the same redshift. This is consistent with Gamma-Ray Bursts exploding in inner regions of their host galaxies, while quasar absorbers probe peripheral regions of the intervening galaxies where the sky cross-section is the largest \citep{Prochaska07, Fynbo08}.

Figure~\ref{fig:gas} shows the dust surface density as a function of cold gas surface density. The cold gas refers to the sum of neutral atomic, \hi, and molecular, \nmol, except for quasar absorbers where the molecular gas is found to be negligible \citep{ledoux2003}.  
The grey lines show constant dust-to-gas ratios in steps of 1 dex. The dust column densities roughly follow the total hydrogen column densities, though with a 4-order of magnitude spread at a given $\Sigma_{\rm Gas}$. \cite{RomanDuval14} used resolved {\it Herschel} infra-red maps of Magellanic clouds in combination with 21cm, CO and H$\alpha$ observations to infer the relation between the atomic, molecular and ionised gas with dust surface density. The authors report a clear trend of increasing dust surface density with increasing gas surface density in the diffuse interstellar medium, albeit for a medium with consistent metallicity. Our depletion-derived results show a similar scaling in Figure~\ref{fig:gas}. \cite{RomanDuval17} further explore the relation based on IRAS and Planck observations and find an increase by a factor three of the dust-to-gas ratio going from the diffuse to the dense insterstellar medium, in line with elemental depletions results. In the Milky Way \citep{jenkins2009} and Small
Magellanic Cloud \citep{jenkins2017}, the
fraction of metals in the gas phase decreases with increasing
hydrogen volume density and column density, albeit at
different rates for different elements. 
Gamma-Ray Burst host galaxies display low dust surface densities in comparison with quasar absorbers while still bearing large amounts of gas \citep{Jakobsson06, Fynbo09}, even at relatively high gas surface densities. One possible reason for this difference is that Gamma-Ray Burst host galaxies have overall low metallicities and low dust content \citep{Perley16, Kruehler15, Savaglio09}, although dustier systems do exist \citep{Perley11, Perley13}. 
We cannot exclude that systems with high dust surface density (even in the low metallicity surface density regime) are missing from current sample based on UV dust-depletion observations. Interestingly, there is a dearth of systems with high gas surface density and low dust surface density which can not be attributed to an observational bias, because systems lying in this parameter space - if they exist - would have small {extinction}.

The top panel of Figure~\ref{fig:sigvmetal} displays the dust surface density as a function of gas metallicity. 
There is a relatively tight correlation between the dust column densities and the dust-corrected metallicity estimates at all cosmic times. Works from \cite{jenkins2009, RomanDuval22b} report dust-depletion measurements assuming a global metallicity of [X/H]=0 for the Milky Way, [X/H]=$-$0.30 for the Large Magellanic Cloud and [X/H]=$-$0.70 for the Small Magellanic Cloud. Additionally, the top panel of Figure~\ref{fig:sigvmetal} shows a clear correlation between $\Sigma_{\rm dust}$ and metallicity. These results are in line with the dust-to-metal ratio increasing with metallicity \citep{wiseman2017,decia2013,Peroux20}. 
The bottom panels of Figure~\ref{fig:sigvmetal} illustrate the distribution in dust surface density {\it at fixed metallicity} materialised by the {grey shaded areas} in the top panel. In the Large Magellanic Clouds in particular, the dispersion in $\Sigma_{\rm dust}$ is larger than in quasar absorbers for a given metallicity, providing further evidence that the metallicity within the Clouds might vary. {Likewise, we note that the Magellanic Clouds dust distributions are skewed towards higher values than quasar absorbers with the similar metallicity. This effect might be related to (i) quasar absorbers probing random position in the galaxy, and therefore being more likely to probe the outermost parts, and (ii) Magellanic Clouds are infalling satellites, so the gas compression coming from the ram pressure might boost dust formation by increasing the densities. We stress that further differences in $\Sigma_{\rm dust}$ might be not revealed here because of the assumption of fixed metallicity in the Magellanic Clouds.} 

{The right most panel of Figure~\ref{fig:sigvmetal} shows a comparison of quasar absorbers with Gamma-Ray Bursts in the range -1.5$<$[M/H]$<$-0.5 indicating that the latter have larger dust surface density values. Indeed, at a given metallicity, $\Sigma_{\rm dust}$ is higher in systems such as Gamma-Ray Bursts that are closer to the denser and colder parts of their galaxies, where the physical conditions (high density, low temperature and high pressure) favour the formation of molecules \citep{blitz2006} and likely dust. }

\begin{figure*}
	\includegraphics[width=1.8\columnwidth]{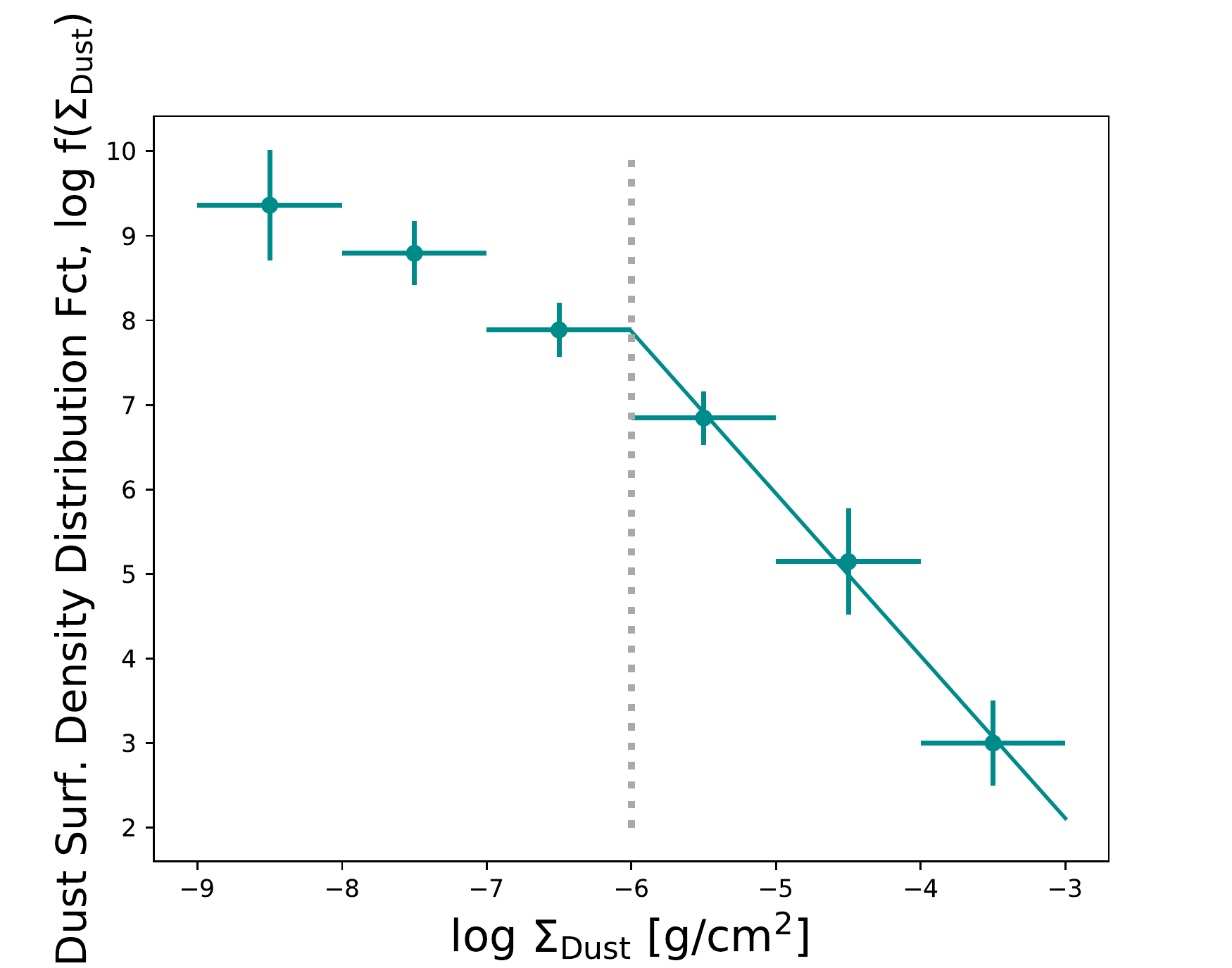}
 	\caption{{\bf Dust surface density distribution function.} This function is computed in analogy with the gas column density distribution function. We note a turn-over at low dust surface densities to the left of the dotted grey line. We interpret this feature as due to incompleteness in the sample. Indeed, the data plotted here are focused on the larger \nhi\ column density quasar absorbers. We fit the high dust surface density values, $\log \Sigma_{\rm Dust}\geq -6$, with a simple power law of the form: $\log f( \Sigma_{\rm Dust})=-1.92 \times \log \Sigma_{\rm Dust} - 3.65$. {This observed relation, which can be computed by spatially resolved simulations predicting dust mass functions through 2D projection, provides new constraints on modern dust models.}
   }
    \label{fig:ColDist}
\end{figure*}

\subsection{Dust Surface Density Distribution Function}

Next, we use these observations to calculate the dust surface density distribution function. To this end, we propose to use an analogue of the gas {(\nhi\ or \nmol)} column density distribution functions \citep{peroux2003, zwaan2005, zwaan2006, klitsch2019, Peroux20, Szakacs22}. We express the function as follows:

\begin{equation}
    f(\Sigma_{\rm Dust})=\frac{\mathcal{N}}{\Delta \Sigma_{\rm Dust} {\rm [g/cm^2]}}
\end{equation}

where $\mathcal{N}$ denotes the number of absorbers in the dust surface density bin $\Delta \Sigma_{\rm Dust}$ \citep[see also][]{churchill2003, richter2011}. Since the function is not normalised by the redshift path, it depends on the number of sightlines in the survey. Our quasar absorber sample comprises a total of 247 systems. The resulting function for quasar absorbers is displayed in Figure~\ref{fig:ColDist}. The data appear to follow a power-law distribution, with the turn-over at small column densities.
This turn-over kicks in at dust surface densities below $\log \Sigma_{\rm Dust} \leq -6$. We interpret this feature as due to incompleteness in the observed sample. We stress that the observations reported here are focused on the larger \nhi\ column density quasar absorbers. It is likely that at low gas surface density, there are a number of lower dust surface density systems which are currently not included in the sample. Indeed, the sample is limited to strong quasar absorbers with \logNHI$>$20.3. { Figure~\ref{fig:gas} shows that quasar absorbers with $\log $N(H) $\leq$ 20.3 (top x-axis) will mostly have $\log \Sigma_{\rm Dust}$ $\leq$ -6 (left y-axis). We note that these systems are not included by construction in the dust surface density distribution function presented here. } 
For this reason, we choose to fit the function without taking the low $\Sigma_{\rm Dust}$ values into account and with a simple power law of the form: 

\begin{equation}
    f(\Sigma_{\rm Dust})= C \Sigma_{\rm Dust}^{-\delta}
\end{equation}

which we rewrite as:
    
\begin{equation}
    \log f(\Sigma_{\rm Dust})= - \delta \times \log \Sigma_{\rm Dust} + \log C \\
  = -1.92 \times \log \Sigma_{\rm Dust} - 3.65
\end{equation}

for surface densities with $\log \Sigma_{\rm Dust} \geq -6$ {in units of g/cm$^2$. Here, $\log f(\Sigma_{\rm Dust})$ is the number of systems with dust surface density $\Sigma_{\rm Dust}$ 
per unit of dust surface density $\Delta \Sigma_{\rm Dust}$. It is therefore expressed in number of systems cm$^2$/g. C is the normalisation factor}. The fit is also shown in Figure~\ref{fig:ColDist}. \cite{Rees88} demonstrated that assuming randomly distributed lines-of-sight through spherical isothermal halos {the column density distribution function}, $f(N)$, was shown to have a power law of slope $\delta$=5/3. Interestingly, \cite{Kim01} report a slope $\delta \sim$1.5 for HI absorbers and \cite{churchill2003, richter2011} measure slopes ranging $\delta$=1.5-2.0 for MgII, FeII, MgI and CaII, respectively. Therefore, the slope of the dust surface density distribution function, $\delta$=1.92{$\pm$0.13}, is {possibly} steeper than for neutral gas and {on the high-end with respected to} metal absorbers. {Finally, we note that this observed relation, which simulations predicting dust mass function  \citep{Pozzi20, Millard20} will be able to compute through 2D projection, provides new constraints on modern dust models \citep{Mckinnon17, graziani2019, li2019, hou2019, Baes20}.}

We caution that these UV-depletion results could potentially be incomplete due to observational biases affecting e.g., the dust surface density distribution. For example, in the Milky Way and the Magellanic Clouds, UV sight-lines are biased toward the less reddened stars/lower surface densities by the sensitivity of the Ultra-Violet telescopes. Similarly, such effects apply to Gamma-Ray Bursts and quasar samples so that the dustier objects might be missed from current sample. Several results \citep{Ellison2009} indicate that these effects are minimum, but one cannot fully exclude that dustier objects exists and have been missed from dust-depletion studies presented here.


\section{Conclusions} \label{sec:conclusions}

In this work we have looked at a observable, namely $\Sigma_{\rm Dust}$,
to put novel constraints on simulations of dust. Indeed, reproducing dust masses over cosmic times requires that dust grow in the interstellar medium, and therefore that the dust
properties change significantly with environment,
particularly density. To this end, we gathered observations from the Milky Way, Large and Small Magellanic Clouds and high-redshift galaxies traced by Gamma-Ray Burst host galaxies and quasar absorbers. By putting all these results together we can make a new appraisal of the dust surface density (dust column density) expressed in g per cm$^2$ or alternatively in M$_{\odot}$/kpc$^2$ across cosmic times measured through dust depletion. We also contrast the observational measurements with recent hydrodynamical simulations. Our main results are:

\begin{itemize}
    \item The dust surface densities increase with cosmic time with a large scatter at any given redshift. 
    
    \item The dust surface densities are also a function of the total gas surface densities in the same systems, with increasing dust surface density increasing with total hydrogen surface density, although the scatter in the relation is 4 orders of magnitude. 
    
    \item There is a tight correlation between the dust column densities and the dust-corrected metallicity estimates at all cosmic times.

    \item We introduce the dust surface density distribution function -- in analogy with the cold gas column density distribution function. We note a turn-over at low dust surface densities. We interpret this feature as due to incompleteness in the sample. We provide {a} fit to the observed distribution of the form: $\log f(\Sigma_{\rm Dust})$ {[number of systems per unit dex]} $=-1.92 \times \log \Sigma_{\rm Dust} - 3.65$ which proves steeper than for neutral gas and metal absorbers. 

\end{itemize}

\section*{Data Availability}

Data directly related to this publication and its figures is available on request from the corresponding author.

\section*{Acknowledgements}
We are grateful to Omima Osman, Enrico Garaldi, Qi Li, Julia Roman-Duval and Sandra Savaglio for helpful comments. {We thank the anonymous referee for their suggestions which improved the results presented here.} This research was supported by the International Space Science Institute (ISSI) in Bern, through ISSI International Team project \#564 (The Cosmic Baryon Cycle from Space). ADC acknowledges support from the Swiss National Science Foundation under grant 185692. JCH recognizes support from the US National Science Foundation through grant AST-1910255.

\bibliographystyle{mnras}
\bibliography{refs}

\newpage
\pagebreak
\appendix
\section{Observed dust properties at various cosmic times.}
\label{app}

Table~\ref{tab:obs} lists the observed physical properties of different environment plotted in Figures~\ref{fig:redshift}, \ref{fig:gas} and \ref{fig:sigvmetal}. We derive these quantities from data of the the Milky Way \citep{DeCia21}, the Large and Small Magellanic Clouds observations \citep{RomanDuval22a}, the Gamma-Ray Bursts information \citep{Bolmer2019} and the quasar absorbers data \citep{Peroux20}. This extract shows the first few entries while the full table is available as machine-readable on-line material.

\begin{table}
\begin{center}
\caption{{\bf Summary of the observational data used in this study.} The first column refers to the systems under study, the second column lists our computation of the logarithm of the dust surface density in units of g per cm$^2$, the third column provides the redshift, while the fourth column tabulates the total gas surface density and the last column provides the dust-depletion corrected total metallicity with respect to solar.} \label{tab:obs}
\begin{tabular}{lcccc}
    \hline
System   &log $\Sigma_{\rm Dust}$ &Redshift &log $\Sigma_{\rm Gas}$ & Metallicity\\
       &[g/cm$^{-2}$] & &[g/cm$^{-2}$] &[M/H]$_{\rm total}$\\
    \hline
Milky Way	     &$-$5.05$\pm$0.61	&0.000 &$-$2.49$\pm$0.20 &$-$0.39$\pm$ 0.18	\\
Milky Way	     &$-$5.04$\pm$0.52	&0.000 &$-$2.31$\pm$0.20 &$-$0.56$\pm$ 0.09	\\
Milky Way	     &$-$5.48$\pm$0.55	&0.000 &$-$3.12$\pm$0.20 &$-$0.18$\pm$ 0.13	\\
Milky Way	     &$-$4.87$\pm$0.56	&0.000 &$-$2.33$\pm$0.20 &$-$0.41$\pm$ 0.13	\\
Milky Way	     &$-$4.81$\pm$0.62	&0.000 &$-$2.12$\pm$0.20 &$-$0.50$\pm$ 0.20	\\
Milky Way	     &$-$5.35$\pm$0.56	&0.000 &$-$3.46$\pm$0.20 &$+$0.28 $\pm$0.13 	\\
...\\
\hline
\end{tabular}
\end{center}
\end{table}

\end{document}